\documentclass[iop,apjl]{emulateapj}

\usepackage{graphicx} 
\usepackage{dcolumn} 
\usepackage{bm} 
\usepackage{amsfonts,amsmath,amssymb,mathrsfs}
\usepackage{color}
\usepackage{epsfig}
\usepackage{natbib,times}
\usepackage{url}
\usepackage{times}

\pdfoutput=1

\shorttitle{Magnetar Magnetic Fields from BNS mergers}
\shortauthors{Giacomazzo et al.}

\begin{document}

\title{Producing Magnetar Magnetic Fields in the Merger of Binary Neutron Stars}

\author{Bruno {Giacomazzo}\altaffilmark{1,2}, Jonathan
  {Zrake}\altaffilmark{3}, Paul C. {Duffell}\altaffilmark{4}, Andrew
  I. {MacFadyen}\altaffilmark{4}, and Rosalba {Perna}\altaffilmark{5}}

\altaffiltext{1}{Physics Department, University of Trento, via
  Sommarive 14, I-38123 Trento, Italy}

\altaffiltext{2}{INFN-TIFPA, Trento Institute for Fundamental Physics
  and Applications, via Sommarive 14, I-38123 Trento, Italy}

\altaffiltext{3}{Kavli Institute for Particle Astrophysics and
  Cosmology, Stanford University, Menlo Park, CA 94025, USA}

\altaffiltext{4}{Center for Cosmology and Particle Physics, Physics
  Department, New York University, New York, NY 10003, USA}

\altaffiltext{5}{Department of Physics and Astronomy, Stony Brook
  University, Stony Brook, NY 11794-3800, USA}

\begin{abstract}
The merger of binary neutron stars (BNSs) can lead to large
amplifications of the magnetic field due to the development of
turbulence and instabilities in the fluid, such as the
Kelvin-Helmholtz shear instability, which drive small-scale dynamo
activity. In order to properly resolve such instabilities and obtain
the correct magnetic field amplification, one would need to employ
resolutions that are currently unfeasible in global general
relativistic magnetohydrodynamic (GRMHD) simulations of BNS
mergers. Here, we present a subgrid model that allows global
simulations to take into account the small-scale amplification of the
magnetic field which is caused by the development of turbulence during BNS
mergers. Assuming dynamo saturation, we show that magnetar-level
fields ($\sim 10^{16}\,{\rm G}$) can be easily reached, and should
therefore be expected from the merger of magnetized BNSs. The total
magnetic energy can reach values up to $\sim 10^{51}\,{\rm erg}$ and
the post-merger remnant can therefore emit strong electromagnetic
signals and possibly produce short gamma-ray bursts.
\end{abstract}


\section{Introduction}
\label{sec:introduction}

Binary neutron star (BNS) mergers are among the most powerful sources
of gravitational waves (GWs) which are expected to be detected in the
next few years by advanced LIGO and Virgo, and they are also the main
candidates for the central engine of short gamma-ray bursts (SGRBs,
e.g., see~\citealt{Berger2013}). The two main scenarios for the
central engine of SGRBs involve the formation of a strongly magnetized
torus around a spinning black hole (e.g.,
\citealt{2011ApJ...732L...6R,Giacomazzo2013} and references therein)
or the formation of a long-lived magnetar (e.g.,
\citealt{Rowlinson2013,Giacomazzo2013magnetar}). Neutron stars are
often magnetized and, during the merger of two NSs, magnetic fields
can be strongly amplified via instabilities in the plasma, such as the
Kelvin-Helmholtz (KH) instability~\citep{Price2006, Anderson08,
  Baiotti08, Giacomazzo:2009mp, Giacomazzo2011}. While previous
Newtonian simulations have shown that magnetic fields can be amplified
by several orders of magnitude, reaching magnetar-level fields of $\sim
10^{15}\, {\rm G}$ when starting with $\sim 10^{12}\, {\rm G}$
\citep{Price2006}, subsequent and independent studies in full general
relativistic magnetohydrodynamics (GRMHD) did not find such large
amplifications~\citep{Giacomazzo:2009mp, Giacomazzo2011,
  Kiuchi2014}. GRMHD simulations used grid-based codes with
  resolutions that were not sufficiently high to resolve the smallest
  turbulent scale and reach convergence. For example, in the presence
  of KH instabilities, higher resolutions can resolve
  smaller-scale vortices~\citep{Baiotti08} and this can lead to
  stronger magnetic field amplifications. In addition, local
high-resolution simulations performed in the last few years have shown
that magnetic fields can indeed be amplified at merger by several
orders of magnitude and reach equipartition with the kinetic energy of
the turbulent fluid~\citep{Obergaulinger2010,Zrake2013}. Unfortunately,
in order to fully resolve such scales, one would need to employ
resolutions of the order of $\sim 0.1\, {\rm m}$ or
higher~\citep{Obergaulinger2010,Zrake2013}, which is currently
impossible to obtain in global simulations of BNS mergers (where
resolutions are of the order of $\sim 100 \, {\rm m}$).

Here, we develop a subgrid model which allows global GRMHD
simulations of BNS merger to include small-scale effects and, in
particular, properly resolve the magnetic field amplification during
merger. Other subgrid models have already been used with success in
other scenarios, such as magnetic field amplification in accretion
disks~\citep{Sadowski2014}, and here we show for the first time their
useful role during BNS mergers.

Section~\ref{sec:methods} details our numerical methods and the
initial models, and in section~\ref{sec:subgrid} we provide a
detailed description of the implementation of the subgrid
model. Section~\ref{sec:dynamics} describes the evolution of the
magnetic field when the subgrid model is implemented. In
section~\ref{sec:global} we discuss whether this amplification is only
localized in a small region of the domain or if it also happens on a
global scale, while in section~\ref{sec:energy} we discuss the
magnetic energy generated by the turbulence and its impact on
electromagnetic emission, including the possible production of
SGRBs. Finally, section~\ref{sec:conclusions} summarizes our main
results. For convenience, we use a system of units in which
$c=G=M_\odot=1$, unless explicitly stated otherwise.

\begin{figure*}[t]
  \centering
  \begin{tabular}{cc}
    \includegraphics[width=.4\textwidth]{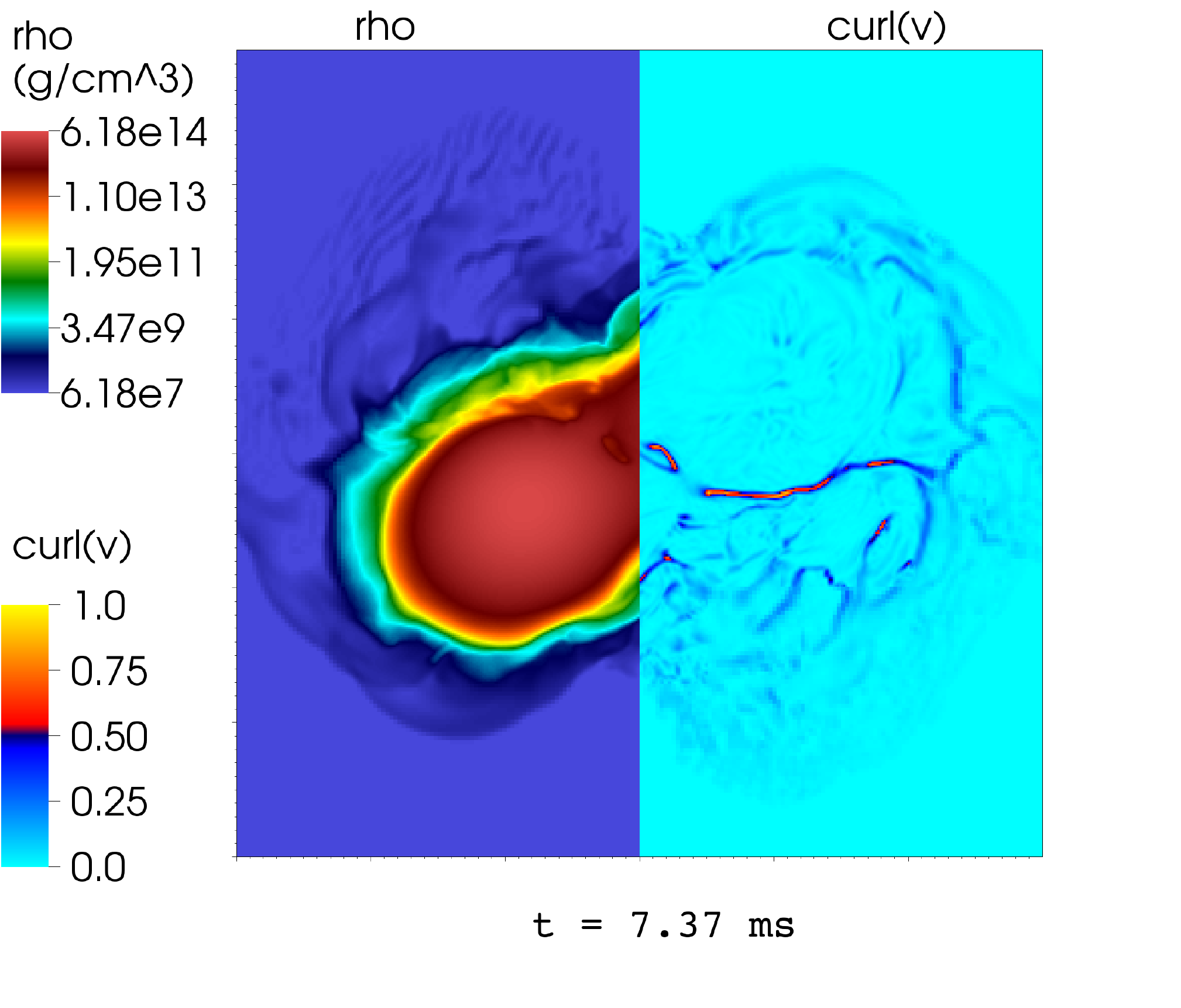} &
    \includegraphics[width=.4\textwidth]{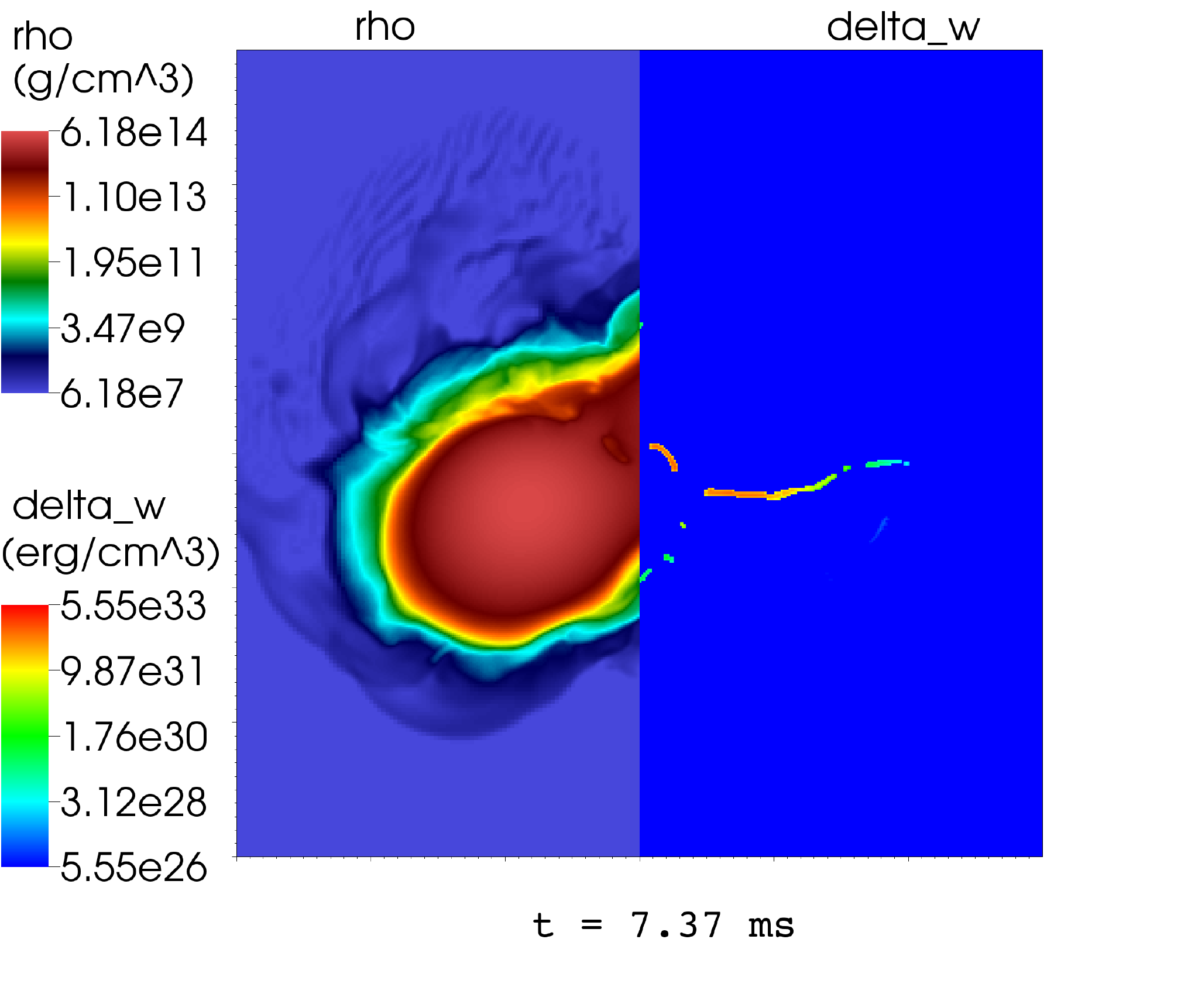} \\
    \includegraphics[width=.4\textwidth]{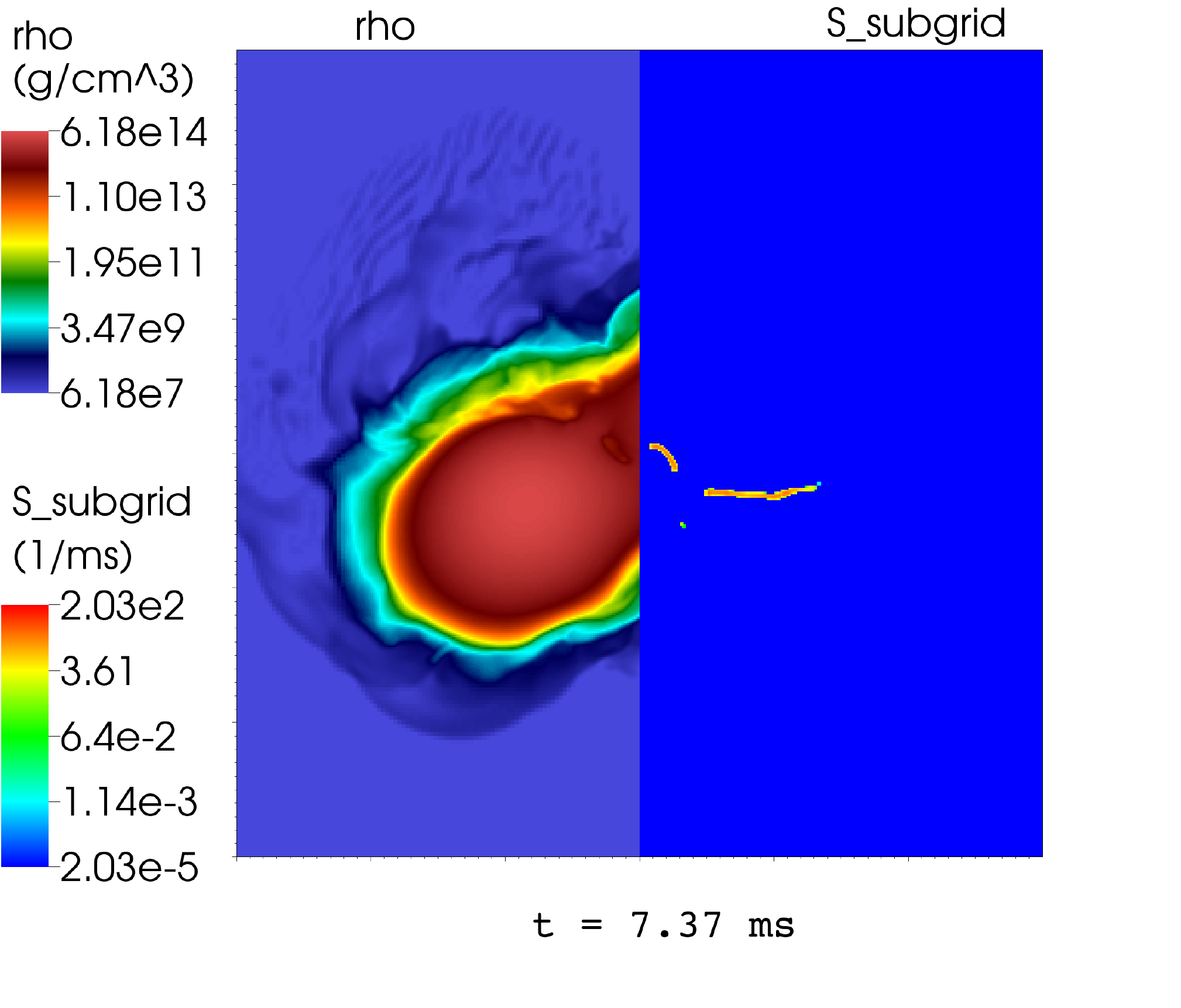} &
    \includegraphics[width=.4\textwidth]{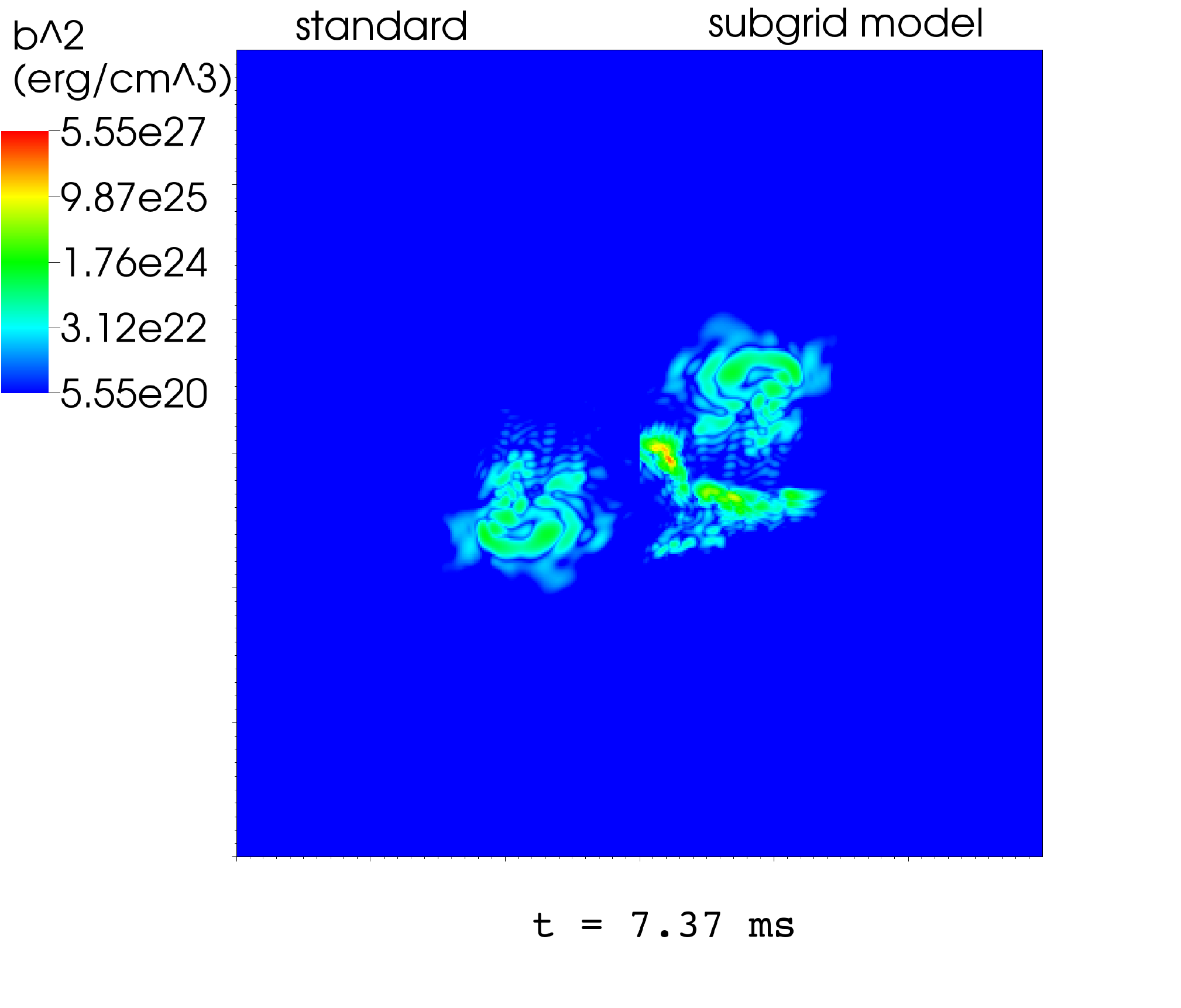}
  \end{tabular}
  \caption{First three panels show on the left side the value on
    the $xy$ equatorial plane of the rest-mass density $\rho$ (in
    ${\rm g}\, {\rm cm}^{-3}$) and on the right side the value of
    $\left|{\nabla} \times {v}\right|$ in geometric units
    (top left panel), of $\Delta w$ in ${\rm erg}\, {\rm cm}^{-3}$
    (top right panel), and of $S_{\rm subgrid}$ in ${\rm ms}^{-1}$
    (bottom left panel). They all refer to our fiducial run with the
    subgrid model. The subgrid model in our runs is applied in those
    regions in which the vorticity $\left|{\nabla} \times
    {v}\right|$ is larger than $c_3=0.5$ (in geometric units) and
    $\Delta w$ for computational reasons is calculated only in those
    regions. The bottom right panel shows instead the magnetic energy
    density $b^2$ in ${\rm erg}\, {\rm cm}^{-3}$ for a ``standard''
    evolution (left side) and for the case in which the subgrid model
    is implemented (right side). All panels refer to time $t\sim 7.4\,
    {\rm ms}$ and they cover the region between $\sim -45\, {\rm km}$
    and $\sim 45\, {\rm km}$ in both $x$ and $y$. These four panels
    show that our subgrid model amplifies the magnetic field mainly in
    the region near the center (where the KH instability is more
    active, e.g., see~\citealt{Baiotti08}) and excludes low-density
    zones near the artificial atmosphere.\label{figure1}}
\end{figure*}

\section{Numerical Methods and Initial Data}
\label{sec:methods}
The simulations presented here were performed using our fully GRMHD
code \texttt{Whisky}~\citep{Giacomazzo2007, Giacomazzo2011,
  Giacomazzo2013magnetar} which is coupled with the publicly available
Einstein Toolkit~\citep{Loeffler2012}. In particular, the spacetime
evolution is computed using the \texttt{McLachlan}
code~\citep{Loeffler2012} while \texttt{Whisky} solves the equations
of GRMHD written in a conservative form~\citep{Giacomazzo2007}. In
order to guarantee the divergence-free character of the magnetic field,
we directly evolve the vector potential using the modified Lorenz
gauge~\citep{Farris2012, Giacomazzo2013magnetar}.

We also use adaptive mesh refinement via the \texttt{Carpet}
driver~\citep{Loeffler2012} by adopting six refinement levels with the
finest grids completely covering each of the NSs. After the merger, the
finest grid is enlarged in order to cover the hypermassive neutron
star (HMNS; \citealt{Giacomazzo2011}). Our fiducial runs have a
resolution of $\sim 225 \,{\rm m}$ on the finest grid (see discussion
in Sec.~\ref{sec:dynamics}) while the coarsest grid extends up to
$\sim 778 \,{\rm km}$.

The initial data was produced using the publicly available code {\tt
  LORENE}~\citep{Taniguchi02b}. The matter is modeled using a
polytropic equation of state (EOS) $p = K \rho^{\Gamma}$, where $p$ is
the pressure, $\rho$ the rest-mass density, $K=123.6$ and
$\Gamma=2$. An ideal-fluid EOS with $\Gamma=2$ is used during the
evolution in order to allow for shock heating during the merger. The
initial data is the same ``high-mass'' model used in our previous
simulations~\citep{Baiotti08,Giacomazzo2011,2011ApJ...732L...6R} and it consists
of an equal-mass system of two NSs with gravitational
mass\footnote{The gravitational mass is measured when the stars
    are isolated.} $\sim 1.5\,M_\odot$ at an initial coordinate
separation of $\sim 46\, {\rm km}$ (approximately $2.5$ orbits before
merger). The magnetic field is initially purely poloidal and aligned
with the angular momentum of the binary as
in~\citet{Giacomazzo2011}. The initial amplitude of the maximum of the
magnetic field, as measured by a normal observer, is $\sim 2.5
\times 10^{12}\, {\rm G}$.

\section{The Subgrid Model}
\label{sec:subgrid}
Our subgrid model is intended to account for electromotive forces
arising from unresolved fluctuations in the magnetic field and bulk
fluid velocity.
In particular, we assume that the unresolved turbulence, and the
velocity field associated with it, gives rise to an extra electric
field $\vec{E}_{\rm subgrid}$ which is added to the right-hand side of the
evolution equation for the vector potential $\vec{A}$:
\begin{equation}
  \partial_t \vec{A} = -\vec{E}_{\rm ideal} - \vec{E}_{\rm subgrid} \,,
\end{equation}
where $\vec{E}_{\rm ideal}$ is the standard electric field coming
from the ideal-MHD equations and is computed using the flux-CD
approach as described in \citet{Giacomazzo2011}. We note that this new
evolution equation for the vector potential does not violate Maxwell's
equations, but it may not satisfy the ideal-MHD condition. Our
assumption is that $\vec{E}_{\rm subgrid}$ would naturally arise in very
high-resolution simulations where the turbulence is fully resolved,
but that it is currently missing due to the still low resolution employed
in BNS simulations.
We therefore adopt a closure scheme intended to account for
small-scale dynamo action driven by the turbulent cascade at the
unresolved length scales.\footnote{Such a closure scheme is not
  unique and a detailed numerical analysis with local simulations will
  be performed in a future paper in order to further assess the
  robustness of our subgrid model.} The effect of this dynamo has been
found, through detailed local simulations \citep{Zrake2013} of
relativistic MHD turbulence, to drive the large-scale (resolved)
magnetic energy density toward equipartition with the local turbulent
kinetic energy density on a timescale given by the turn over of the
energy containing eddies. We characterize the turbulent kinetic energy
density by the field $\Delta w$, whose value is obtained through a
prescription outlined below. The exponentially growing solution is
then given by
\begin{equation}
  \vec{E}_{\rm subgrid} = -S_{\rm subgrid} \vec{A}
\end{equation}
and is parameterized around the exponentiation rate $S_{\rm subgrid}$ which we
take to be a fraction $c_1$ of the local fluid vorticity. Exponentiation of the
magnetic field stops when the local electromagnetic energy density is a fraction
$c_2$ of $\Delta w$. Thus, our closure scheme is parameterized around the
following prescription for $S_{\rm subgrid}$:
\begin{eqnarray}
  S_{\rm subgrid} &\equiv & c_1 \max\left(\left|{\nabla} \times {v}\right|-c_3,0\right) \times \nonumber\\
                      & & \max\left(1-c_4\frac{\rho_{\rm atmo}}{\rho},0\right) \times \nonumber\\
                      & & \max\left(1-\frac{b^2}{c_2\Delta w},0\right)\,, \label{eq_subgrid}
\end{eqnarray}
where $\rho_{\rm atmo}=10^{-10}$ is the value of the rest-mass density
$\rho$ in the artificial atmosphere, $b^2$ is the magnetic energy
density, and $\Delta w$ is a measure of the kinetic energy of the
turbulent portion of the fluid~\citep{Duffel2013}. The coefficients
$c_1=0.5$, $c_2=0.6$, $c_3=0.5$, and $c_4=10^4$ have been chosen based on
the results of~\citet{Zrake2013} (coefficients $c_1$ and $c_2$) and in
order to avoid spurious magnetic field amplifications (coefficients
$c_3$ and $c_4$), especially in regions near the surface of the neutron
star where the (flat-space) vorticity, $|{\nabla} \times
{v}|$, is artificially high because of a jump in the value of the
velocity (since ${v}$ is set to zero in the artificial
atmosphere). We note, in particular, that the value of $c_2$ sets the
saturation level for the magnetic field and our value ($c_2=0.6$)
stops the magnetic field growth when the magnetic energy density is
equal to $60\%$ of the kinetic energy density, in agreement with the
results of~\citet{Zrake2013}. The coefficient $c_1$ was chosen
instead to allow the growth to happen on a timescale of $\sim 1\,{\rm
  ms}$~\citep{Zrake2013}. Note that the coefficients $c_3$ and $c_4$
instead need to be fine tuned in order to avoid spurious amplifications,
especially at the NS surface.

\begin{figure}[t]
  \centering
    \includegraphics[width=.4\textwidth]{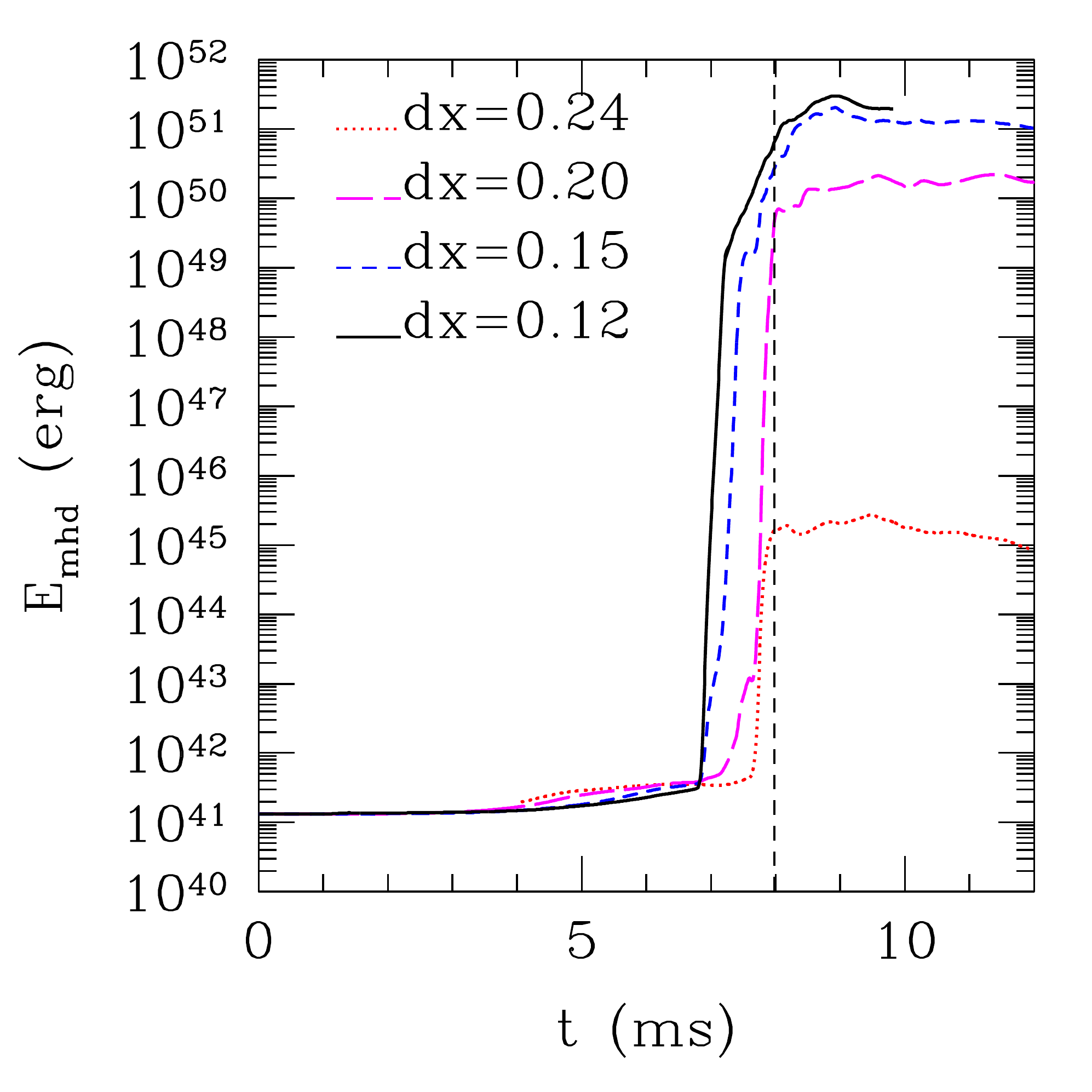}
  \caption{Evolution of the total magnetic energy when the subgrid
    model is implemented and for different resolutions: the solid black
    line refers to our highest-resolution run with $\Delta
    x=0.12\approx 180{\rm m}$, the blue dashed line to $\Delta
    x=0.15\approx 220{\rm m}$, the purple long dashed line to $\Delta
    x=0.20\approx 300{\rm m}$, and the red dotted line to $\Delta
    x=0.24\approx 360{\rm m}$ (all resolutions $\Delta x$ refer to the
    resolution on the finest grid, i.e., the one covering the
    NSs). The vertical black dashed line indicates the time of the
    merger of the two NS cores (which also corresponds to the maximum
    amplitude of the gravitational wave signal), while the external
    layers start merging a few ms before. \label{figure2}}
\end{figure}

$\Delta w$ is computed following~\citet{Duffel2013} where for an
  ideal-fluid EOS we have $w\equiv\rho+p/(\Gamma-1)$, and hence (since
  $\Gamma=2$ in our case)
\begin{equation}
  \Delta w = (<\rho>_{\rm Cons} + <p>_{\rm Cons}) - (<\rho>_{\rm Vol} + <p>_{\rm Vol})\,,
\end{equation}
where $<\rho>_{\rm Vol}$ and $<p>_{\rm Vol}$ are the simple volume
averages of the rest-mass density and pressure, respectively, while
$<\rho>_{\rm Cons}$ and $<p>_{\rm Cons}$ are the ``conservative''
averages of the rest-mass density and pressure
(see~\citealt{Duffel2013} for details). The last two are computed by
averaging the conserved variables over a region of space and then
computing $\rho$ and $p$ using our standard conservative to primitive
solver~\citep{Giacomazzo2007}. When computing the volume average and
the conservative average of $\rho$ and $p$ one needs to choose the
size of the volume over which the average is computed. In our
simulations, we used a cubic box of size $6 \Delta x$, with $\Delta x$
being the resolution of the grid. In this way, when computing $\Delta
w$ on each grid point, we simply need to know the values of the primitive
and conservative variables on the next three grid points in each
direction (e.g., the three grid points on the left and the three on the
right along the $x$, $y$, and $z$ directions). This choice has been made in
order to reduce the computational cost (i.e., MPI calls) of such
computations.

\section{Magnetic field amplification and saturation}
\label{sec:dynamics}

We ran two simulations using the same initial data, but in one case we
used the subgrid model described in the previous section, while in the
other we evolved the system using the ``standard'' vector potential
equations, i.e., without adding the subgrid term $S_{\rm subgrid}A_i$
to the right-hand side of the evolution equations for $A_i$. Here, we
first report on the results obtained with the new subgrid model and in
the next section we compare with the ``standard'' evolution.

In figure~\ref{figure1} we show, at $t\sim 7.4\, {\rm ms}$, the value
on the equatorial plane of the rest-mass density $\rho$, of
$\left|{\nabla} \times {v}\right|$ (top-left panel), of
$\Delta w$ (top-right panel), of $S_{\rm subgrid}$ (bottom-left
panel), and of the magnetic energy density $b^2$ (bottom-right
panel). In the last panel, in particular, we compare the magnetic energy
density between a ``standard'' evolution (left side) and the case in
which the subgrid model is implemented (right side). From these
figures one can see that the regions where $S_{\rm subgrid}$ is non
zero and the magnetic field is amplified are indeed those where the
KH instability is more active (compare also
with~\citealt{Price2006} and \citealt{Baiotti08}). Indeed, in those regions,
both the vorticity ($|{\nabla} \times {v}|$) and
$\Delta w$ are much larger than zero and have their maximum
values. Note also that the vorticity is also quite large in regions
outside the central region. The choice of our parameters in
equation~\ref{eq_subgrid} is such that those regions are excluded,
since the turbulence there, which is anyway smaller than in the
central regions, is due to the interaction with the artificial
atmosphere.

\begin{figure}[t]
  \centering
    \includegraphics[width=.4\textwidth]{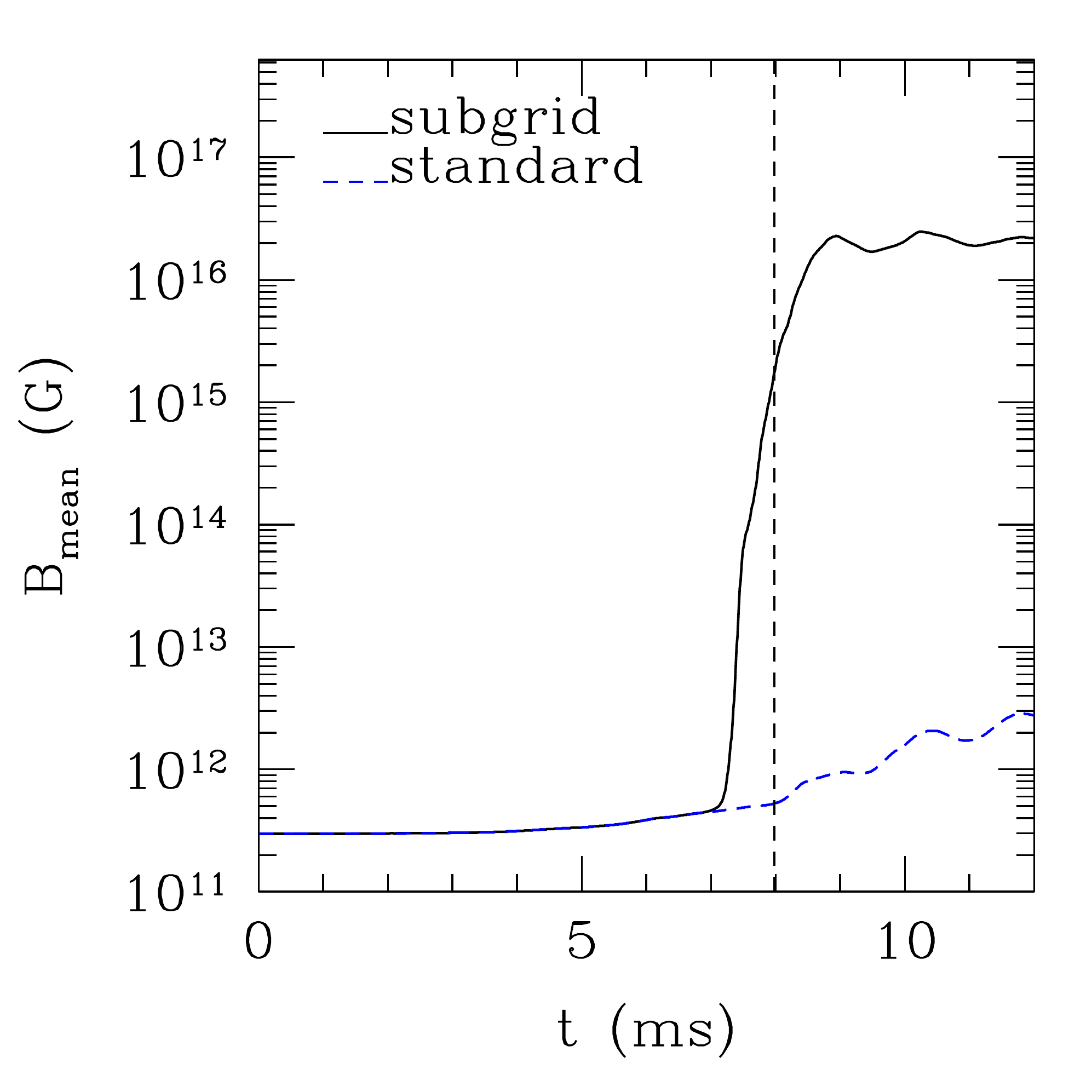}
  \caption{Evolution of the mean value of the magnetic field when the
    subgrid model is implemented (black solid line) and when it is not
    (blue dashed line). The vertical dashed line shows the time of
    merger (when the NS cores collide). While in a ``standard''
    simulation, i.e. a simulation where the subgrid model is not
    implemented, the magnetic field grows by only $\sim 1$ order of
    magnitude, in the simulation implementing the subgrid model the
    magnetic field grows up to $\sim 10^{16}$G and it saturates when
    reaching equipartition with the kinetic energy of the fluid in the
    turbulent regions.\label{figure3}}
\end{figure}

In figure~\ref{figure2}, we show how the amplification changes with
resolution. We reran the same model with one higher resolution
($\Delta x=0.12\approx 180{\rm m}$) and two lower resolutions ($\Delta
x=0.20\approx 300{\rm m}$ and $\Delta x=0.24\approx 360{\rm m}$). In
figure~\ref{figure2}, we plot the evolution of the magnetic energy and,
while the lowest-resolution run (red dotted line) shows only a modest
increase due to just two orders of magnitude amplification in the
magnetic field, the other three resolutions show a much larger
increase. In particular, the two highest-resolution runs produce the
same magnetic energy (and the same magnetic field values), indicating
that saturation has been reached. We note that this is the first time
that such a saturation level is reached in a BNS simulation. Previous
GRMHD simulations were not able to amplify the magnetic field more
than $\sim 1$ order of magnitude at merger and only the Newtonian
simulations by~\citet{Price2006} showed large magnetic field
amplifications, but no saturation was reached and different values
were obtained for different resolutions.

\begin{figure}[t]
  \centering
    \includegraphics[width=.4\textwidth]{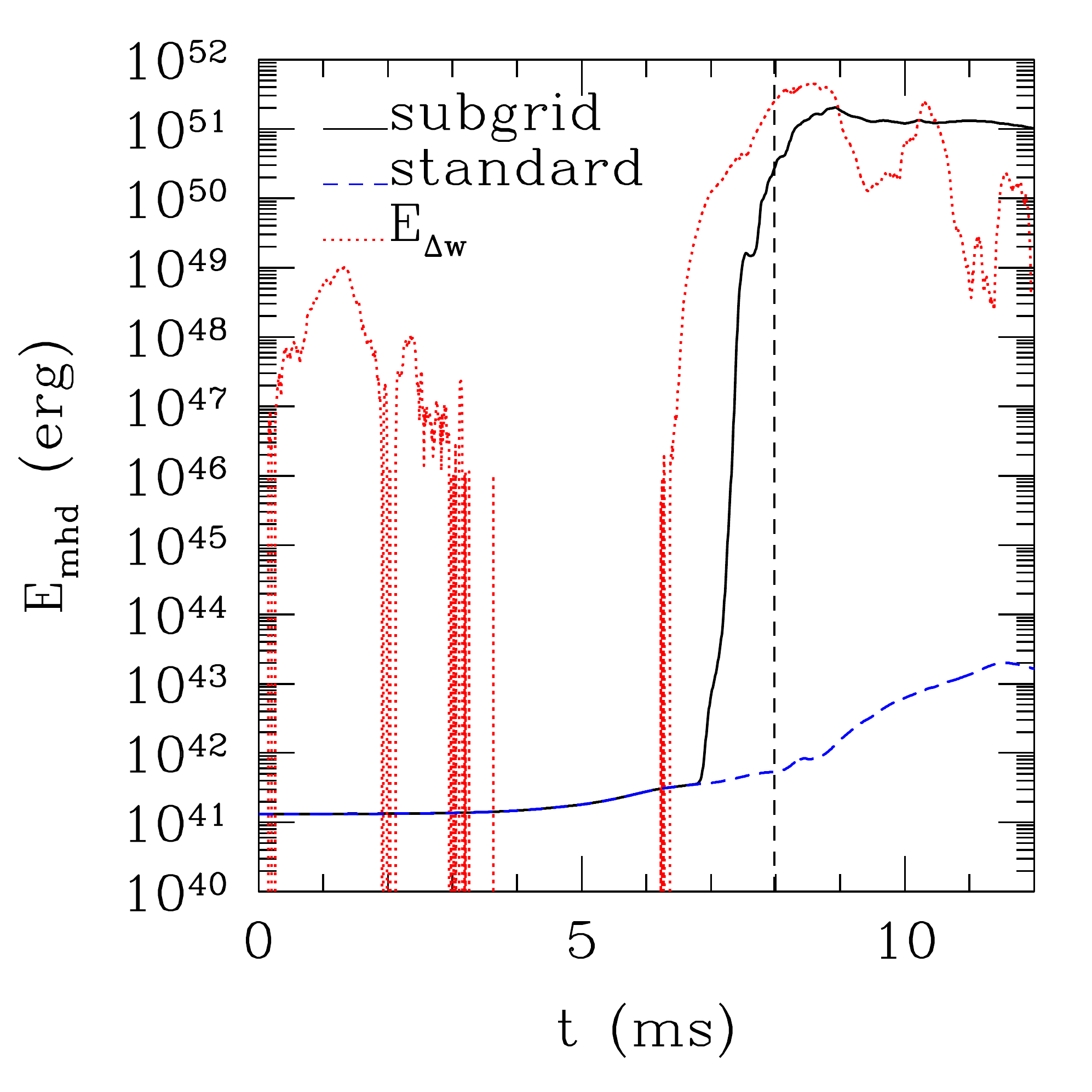}
  \caption{Evolution of the magnetic energy when the subgrid model is
    implemented (black solid line) and when it is not (blue dashed
    line). The vertical black dashed line represents the time of
    merger of the two NS cores. The red dotted line represents instead
    the integral of $\Delta w$ computed where $S_{subgrid}>0$. The
    values of $E_{\Delta w}$ at $t<4 {\rm ms}$ are due to the
    artificial shocks that develop on the NS surfaces during the
    inspiral (due to the fact that we evolve our NSs using an
    ideal-fluid EOS and that our NSs do not have a solid crust). As
    one can easily see, the values of $E_{\Delta w}$ during the first
    part of the inspiral are at least $\sim 2$ orders of magnitude
    below those reached during merger. Moreover, they do not affect
    the evolution of the magnetic field as one can see both from this
    figure (the magnetic energy is constant as in the standard case)
    and from figure~\ref{figure3}, where the mean value, as well as
    the maximum (not shown), of the magnetic field does not grow
    during the inspiral and it is identical to the value in the
    standard run (i.e, when the subgrid model is not
    used). \label{figure4}}
\end{figure}

\section{Local or Global Magnetic Field Amplification?}
\label{sec:global}

In figure~\ref{figure3} we plot the weighted-average of the magnetic field amplitude:
\begin{equation}
  B_{\rm mean} \equiv \frac{\int \rho B dV}{\int \rho dV}\,,
\end{equation}
with $dV$ being the proper volume. The black solid lines represent the
evolution of $B_{\rm mean}$ when the subgrid model is used, while the
blue dashed line represents the ``standard'' evolution. In both cases, we used our
fiducial resolution ($\Delta x=0.15\approx 220{\rm m}$). First of all,
while the maximum of the magnetic field saturates to $\sim
10^{17}\,{\rm G}$ when the subgrid model is used, its mean value
saturates to $\sim 10^{16}\,{\rm G}$. This is a clear indication that
during the evolution the strong magnetic field generated in the
turbulent regions expands and covers a large portion of the HMNS
formed after the merger. The magnetic field amplification is therefore
not killed during the merger but survives and may considerably
affect the post-merger evolution~\citep{Giacomazzo2011}. The blue
dashed line represents instead the mean value of the magnetic field
when the subgrid model is not used. In this case, the magnetic field
grows only by one order of magnitude, as seen in previous
simulations~\citep{Giacomazzo:2009mp, Giacomazzo2011,
  2011ApJ...732L...6R, Kiuchi2014}. By properly taking into account
the amplifications due to the subgrid-scale turbulence, the magnetic
field is amplified by $\sim 4$ orders of magnitude with respect to
what can be afforded by current resolutions. Indeed, we expect that
even without our subgrid model, one should be able to obtain such large
fields when employing sufficiently large resolutions in order to reach
saturation (which may not happen for $\Delta x \gtrsim 0.1\,{\rm m}$).

\section{Magnetic Energy and Gamma-Ray Burst Engine}
\label{sec:energy}

In figure~\ref{figure4} we plot the evolution of the magnetic energy
both when the subgrid model is used (solid black line) and when it is
not (dashed blue line). We also plot the turbulent energy $E_{\Delta
  w}$, i.e., the integral of $\Delta w$ for the case in which the
subgrid model is used. During the merger, $E_{\Delta w}$ grows by
several orders of magnitude and it is followed by an increase in the
magnetic energy. The small delay (less than a ms) between the onset of
the two growths is due to the time the magnetic field needs in order
to be amplified over a large region (the maximum of the magnetic field
grows indeed much earlier and almost simultaneously with the growth of
$E_{\Delta w}$). These results predict the rapid production of a
magnetic energy reservoir on the order of $\sim 10^{51} \, {\rm erg}$
in the moments after the merger onset. It was pointed out
in~\citet{Zrake2013} that a powerful electromagnetic transient could
be powered by the conversion of even a small fraction of that magnetic
energy into escaped photons. There, it was proposed that strong
magnetic fields undergoing violent reconnection in the merger
remnant's atmosphere could power electromagnetic explosions capable of
accelerating plasma to large Lorentz factors. High-energy photons
escape to the observer once the accelerated plasma has overtaken the
slow ($\sim 0.1c$) baryonic outflow. The production of magnetic
explosions is proposed to operate in a manner not unlike those thought
to power the so-called giant magnetar
flares~\citep[e.g.,][]{Thompson1995}. These high-energy transients
could be associated with SGRB precursor emission~\citep{Troja2010} or
even the SGRB itself.

\section{Summary}
\label{sec:conclusions}
We presented the first subgrid model for GRMHD simulations of BNS
mergers that can be used to study small-scale magnetic field
amplifications in global BNS simulations. We show for the first time
that, by assuming dynamo saturation, magnetic field values of the
order of $\sim 10^{16}\, {\rm G}$ can be easily produced in BNS
mergers. This large magnetic field corresponds to a magnetic energy in
the plasma of $\sim 10^{51} \, {\rm erg}$. Such large magnetic fields
and energy can lead to the production of strong electromagnetic
signals~\citep{Siegel2014},
SGRBs~\citep{2011ApJ...732L...6R,Giacomazzo2013magnetar}, and
  also long-lasting GW emission if a stable magnetar is formed
  after merger~\citep{DallOsso2015}. In future papers, we will study
the effect of this field amplification on the evolution of the HMNS,
on the shape of the GW signal, on the possible emission of
relativistic jets, and on ejecta masses and distribution, which may
have consequences for r-process nucleosynthesis as well as jet
collimation and the production of electromagnetic transients from
circum-merger interactions~\citep{2011Natur.478...82N,
  2012ApJ...746...48M}.

As a final comment, we also note that this subgrid model should be
  implemented only as long as GRMHD simulations of BNSs are not able
  to use the resolutions required to fully resolve the
  turbulence. This will require high-order numerical codes combined
  with very high resolutions ($\sim 0.1\,{\rm m}$) which are currently
  unfeasible.

\acknowledgments We acknowledge Zachariah Etienne and an anonymous referee for useful
  comments and suggestions. B.G. and R.P. acknowledge support from NSF
grant No. AST 1009396 and NASA grant No. NNX12AO67G. B.G. also acknowledges
support from MIUR FIR grant No. RBFR13QJYF. This research was also
supported in part by NASA through Fermi grant NNX13AO93G and by
the NSF through grant AST-1009863. This work used XSEDE (allocation
TG-PHY110027) which is supported by NSF grant No. OCI-1053575.


\bibliographystyle{apj}

\begin{thebibliography}{22}

\bibitem[Anderson et al.(2008)]{Anderson08} Anderson, M., 
Hirschmann, E.~W., Lehner, L., et al.\ 2008, Physical Review Letters, 100, 
191101 

\bibitem[{{Baiotti} {et~al.}(2008){Baiotti}, {Giacomazzo}, \&
  {Rezzolla}}]{Baiotti08}
{Baiotti}, L., {Giacomazzo}, B., \& {Rezzolla}, L. 2008, \prd, 78, 084033


\bibitem[Berger(2014)]{Berger2013} Berger, E.\ 2014, \araa, 52, 43

\bibitem[Dall'Osso et al.(2015)]{DallOsso2015} Dall'Osso, S., 
Giacomazzo, B., Perna, R., \& Stella, L.\ 2015, \apj, 798, 25 

\bibitem[{{Duffell} \& {MacFadyen}(2013)}]{Duffel2013}
{Duffell}, P.~C., \& {MacFadyen}, A.~I. 2013, \apj, 775, 87

\bibitem[{Farris {et~al.}(2012)Farris, Gold, Paschalidis, Etienne, \&
  Shapiro}]{Farris2012}
Farris, B.~D., Gold, R., Paschalidis, V., Etienne, Z.~B., \& Shapiro, S.~L.
  2012, Phys. Rev. Lett., 109, 221102

\bibitem[{{Giacomazzo} \& {Perna}(2013)}]{Giacomazzo2013magnetar}
{Giacomazzo}, B., \& {Perna}, R. 2013, \apjl, 771, L26

\bibitem[{{Giacomazzo} {et~al.}(2013){Giacomazzo}, {Perna}, {Rezzolla},
  {Troja}, \& {Lazzati}}]{Giacomazzo2013}
{Giacomazzo}, B., {Perna}, R., {Rezzolla}, L., {Troja}, E., \& {Lazzati}, D.
  2013, \apjl, 762, L18

\bibitem[{Giacomazzo \& Rezzolla(2007)}]{Giacomazzo2007}
Giacomazzo, B., \& Rezzolla, L. 2007, Classical Quantum Gravity, 24, S235

\bibitem[{{Giacomazzo} {et~al.}(2009){Giacomazzo}, {Rezzolla}, \&
  {Baiotti}}]{Giacomazzo:2009mp}
{Giacomazzo}, B., {Rezzolla}, L., \& {Baiotti}, L. 2009, \mnras, 399, L164

\bibitem[{Giacomazzo {et~al.}(2011)Giacomazzo, Rezzolla, \&
  Baiotti}]{Giacomazzo2011}
Giacomazzo, B., Rezzolla, L., \& Baiotti, L. 2011, \prd, 83, 044014


\bibitem[Kiuchi et al.(2014)]{Kiuchi2014} Kiuchi, K., Kyutoku, K., 
Sekiguchi, Y., Shibata, M., \& Wada, T.\ 2014, \prd, 90, 041502 


\bibitem[{{L{\"o}ffler} {et~al.}(2012){L{\"o}ffler}, {Faber}, {Bentivegna},
  {Bode}, {Diener}, {Haas}, {Hinder}, {Mundim}, {Ott}, {Schnetter}, {Allen},
  {Campanelli}, \& {Laguna}}]{Loeffler2012}
{L{\"o}ffler}, F., {Faber}, J., {Bentivegna}, E., {et~al.} 2012, Classical and
  Quantum Gravity, 29, 115001

\bibitem[Metzger \& Berger(2012)]{2012ApJ...746...48M} Metzger, B.~D.,
  \& Berger, E.\ 2012, \apj, 746, 48

\bibitem[Nakar \& Piran(2011)]{2011Natur.478...82N} Nakar, E., \&
  Piran, T.\ 2011, \nat, 478, 82

\bibitem[{{Obergaulinger} {et~al.}(2010){Obergaulinger}, {Aloy}, \&
  {M{\"u}ller}}]{Obergaulinger2010}
{Obergaulinger}, M., {Aloy}, M.~A., \& {M{\"u}ller}, E. 2010, \aap, 515, A30

\bibitem[{Price \& Rosswog(2006)}]{Price2006}
Price, D.~J., \& Rosswog, S. 2006, Science, 312, 719

\bibitem[{{Rezzolla} {et~al.}(2011){Rezzolla}, {Giacomazzo}, {Baiotti},
  {Granot}, {Kouveliotou}, \& {Aloy}}]{2011ApJ...732L...6R}
{Rezzolla}, L., {Giacomazzo}, B., {Baiotti}, L., {et~al.} 2011, \apjl, 732, L6

\bibitem[{{Rowlinson} {et~al.}(2013){Rowlinson}, {O'Brien}, {Metzger},
  {Tanvir}, \& {Levan}}]{Rowlinson2013}
{Rowlinson}, A., {O'Brien}, P.~T., {Metzger}, B.~D., {Tanvir}, N.~R., \&
  {Levan}, A.~J. 2013, \mnras, 430, 1061


\bibitem[S{\c a}dowski et al.(2015)]{Sadowski2014} S{\c a}dowski, 
A., Narayan, R., Tchekhovskoy, A., et al.\ 2015, \mnras, 447, 49 


\bibitem[{{Siegel} {et~al.}(2014){Siegel}, {Ciolfi}, \&
  {Rezzolla}}]{Siegel2014}
{Siegel}, D.~M., {Ciolfi}, R., \& {Rezzolla}, L. 2014, \apjl, 785, L6

\bibitem[{Taniguchi \& Gourgoulhon(2002)}]{Taniguchi02b}
Taniguchi, K., \& Gourgoulhon, E. 2002, Phys. Rev. D, 66, 104019

\bibitem[{Thompson \& Duncan(1995)}]{Thompson1995}
Thompson, C., \& Duncan, R.~C. 1995, Monthly Notices of the Royal Astronomical
  Society, 275, 255

\bibitem[{Troja {et~al.}(2010)Troja, Rosswog, \& Gehrels}]{Troja2010}
Troja, E., Rosswog, S., \& Gehrels, N. 2010, The Astrophysical Journal, 723,
  1711

\bibitem[{{Zrake} \& {MacFadyen}(2013)}]{Zrake2013}
{Zrake}, J., \& {MacFadyen}, A.~I. 2013, \apjl, 769, L29

\end{thebibliography}

\end{document}